\documentstyle[prl,aps,graphicx]{revtex}

\begin{document}
\draft

\wideabs{
\title{Periodic Oscillations of Josephson-Vortex Flow Resistance in
Bi${}_2$Sr${}_2$CaCu${}_2$O${}_{8+y}$}

\author{S. Ooi, T. Mochiku, and K. Hirata}
\address{National Institute for Materials Science,
Sengen 1-2-1, Tsukuba 305-0047, Japan}

\date{\today}
\maketitle
\begin{abstract}
To study the Josephson-vortex system in the intrinsic Josephson
junctions, we have measured the flow resistance as a function of magnetic
field parallel to the $ab$ plane 
in Bi${}_2$Sr${}_2$CaCu${}_2$O${}_{8+y}$ single crystals.
Although it was known that 
the flow resistance increases smoothly with increasing field,
we have found novel oscillations of vortex flow resistance
in the small current regime.
The period of the oscillations corresponds to the field which
is needed to add ``{\it one}'' vortex quantum 
per ``{\it two}'' Josephson junctions.
Commensurability between the lattice spacing of Josephson vortices 
along the $ab$ plane and the size of the junction
is related to the flow velocity of vortices.
The results show that 
Josephson vortices form triangular lattice in the state
where the oscillations occur.

\end{abstract}


}

When an external magnetic field is applied 
along the plane of Josephson junctions in the superconducting state,
the field penetrates into the junctions as quantized Josephson vortices.
Since high-$T_{\rm c}$ cuprates such as Bi$_2$Sr$_2$CaCu$_2$O$_{8+y}$
(BSCCO) naturally consist of atomic-scale Josephson junctions 
 (intrinsic Josephson junctions; IJJ's)
based on the layered crystal structures and large anisotropy \cite{Kleiner92},
the magnetic field penetrated along superconducting planes
forms a Josephson-vortex system.
Recently, the Josephson-vortex system in high-$T_c$ cuprates
has attracted much attention because of fundamental interests
and possibilities for applications.

When the intrinsic pinning by the superconducting layers \cite{Tachiki89} 
becomes strong enough,
the commensuration of the vortex lattice 
with the underlying layered structure becomes important,
so that the drastic changes in the vortex phase diagram is expected.
To clarify thermodynamic phases of the Josephson-vortex system,
both theoretical and experimental 
studies have been done extensively.
Theoretically,
several new phases \cite{Ivlev90}, 
a smectic phase \cite{Balents95}, 
and a tricritical point 
where the order of the melting transition changes \cite{Hu2000}
have been predicted.
Experimentally, 
the important facts such as 
the melting transition of the Josephson-vortex systems 
\cite{Kwok94,Schilling97,Ishida97}, 
the oscillatory melting temperature \cite{Gordeev2000},
and the existence of the vortex smectic phase \cite{Gordeev2000}
have been found in YBa$_2$Cu$_3$O$_{7-\delta}$ (YBCO).

Interesting phenomena appear 
when the vortex dynamics 
is influenced by a periodic one-dimensional pinning potential.
Periodic oscillations of pinning strength 
as a function of the magnetic field $H$
have been observed in an artificially structured superconducting film 
with channels of weak pinning 
in a strong-pinning environment \cite{Pruymboom88}.
The commensurability between vortex lattice and the channel width 
is related to this phenomenon.
Additionally, Gurevich {\it et al.} have reported
the periodic oscillations of in-plane resistance $R(H)$ 
in superconducting multilayers of NbTi/Cu
when the Josephson vortices move perpendicular to the plane \cite{Gurevich96}.
They concluded that the origin is a dynamic matching effect in terms of 
the generation of standing electromagnetic waves in moving vortex lattice
along the $c$-axis.

Although the influence of the periodic pinning potential is expected 
to be more important in highly anisotropic materials such as BSCCO
than in YBCO and others,
there is little information on the Josephson vortices in BSCCO.
In this Letter, 
we report on novel periodic oscillations of the flow resistance
of Josephson vortices as a function of the in-plane magnetic field.
The period of the oscillations suggests that 
the ground state of Josephson vortices is in triangular lattice
when the oscillations occur.

High quality single crystals of BSCCO were grown 
by traveling-solvent floating-zone technique \cite{Mochiku97}. 
A platelet of single crystals was carefully cut into narrow strips. 
After four contacts with electrodes were formed by using a silver paste, 
the center of strips of the single crystals was cut by focused ion beam (FIB). 
The same process to fabricate IJJ's has been used 
in the case of BSCCO whiskers \cite{Kim99}. 
An FIB image of a typical sample is shown in Fig.\ 1.
The inset is the schematic illustration of the junction. 
The dimensions (width ($w$)$\times$length ($l$)$\times$thickness ($t$)) 
of the measured samples A, B and C 
are 7.3$\times$13$\times$1.3, 18$\times$16$\times$1.6 
and 31$\times$17$\times$2.0 $\mu$m$^3$, respectively.
The superconducting transition temperature $T_{\rm c}$ 
is 86 K in all samples.

While resistance of strips was smaller than 1$\Omega$ 
before the fabrication of junctions, 
the resistance of sample A, B and C 
increased to 1400, 560 and 315\ $\Omega$ at room temperature 
after the fabrications, respectively. 
Therefore, the measured resistance is 
almost $c$-axis resistance of the junction parts and
the contribution of in-plane resistivity can be ignored.
In the small mesa crystals of BSCCO,
the flow of Josephson vortices, 
which is not affected very much by pancake vortices,
has been realized 
in the so-called lock-in state \cite{Lee95,Hechtfischer97,Yurgens99}.
The $ab$ plane of the sample was possible 
to orient parallel to the magnetic field to an accuracy of 0.005 degrees
by adjusting the flux-flow resistance as a function of the misorientation angle
to its maximum.


When the field is applied parallel to the $ab$ planes 
in the superconducting state,
the current along the $c$-axis drives Josephson vortices
to the direction perpendicular to both current and field.
The motion of the vortices causes the flow resistance.
Up to the present it has been reported that 
the flow resistance increases smoothly with increasing field
because the number of vortices increases \cite{Hechtfischer97,Yurgens99}.
However, we have discovered
novel oscillations as a function of field as shown in Fig.\ 2.
When the magnetic field is increased,
finite flow resistance appears in 2 $\sim$ 6 kOe
depending on the sample and the current density.
This field may be related to the pinning effect of Josephson vortices
due to the quality of the surface boundary of samples.
The oscillations start from a constant field 
$H_{\rm start} \sim$ 7-8kOe in all samples.
The period $H_{\rm p}$ of the oscillations
is quite constant in the wide range of magnetic fields.
As the magnetic field is increased,
the flow resistance increases accompanying with the oscillations and
suddenly drops above a field $H_{\rm stop}$.

$H_{\rm stop}$ decreases 
with increasing the absolute value of the angle ($\theta$)
between the magnetic field and the $ab$ plane.
This may be due to the breakdown of the lock-in state
because the pancake vortices penetrate into
the junction and are pinned to stop the flow of Josephson vortices
when the $c$-axis component of the magnetic field exceeds 
the $c$-axis lower critical field $H_{\rm c1}$.
Actually, $H_{\rm stop}$ is roughly proportional to 
$1/\sin\theta$ and the coefficient is of the order of $H_{\rm c1}$.
The maximum of $H_{\rm stop}$ could be reached to 40\ kOe
with our alignment set-up.
On the other hand, $H_{\rm start}$ and $H_{\rm p}$ does not almost change
even if the field is intentionally shifted from the optimal position.

The temperature dependence of flow resistance has also been measured
from 4.2\ K to $T_{\rm c}$.
The periodic oscillations can be observed 
at all temperatures below $(T_{\rm c} - 5)$ K.
$H_{\rm p}$ is independent of temperatures.
These facts show that the oscillations relate only to
the field component parallel to the $ab$ plane,
namely, the dynamics of the Josephson vortices.

It is interesting that 
periodic oscillations can be observed 
only in sufficiently small current regime.
$I$-$V$ characteristics of the sample B are shown in Fig.\ 3.
In the fields where the resistance is minimum,
the nonlinearity of $I$-$V$ curves becomes large 
and a kink structure is observed around 50$\mu$A.
The periodic appearance of this structure as a function of fields 
corresponds to the oscillations of flow resistance.
In the larger currents $>$ 200$\mu$A the oscillatory behavior 
disappears thoroughly.

$H_{\rm p}$ depends on the dimensions of the junctions.
To extract $H_{\rm p}$,
the power spectral densities for three samples which are obtained from 
the data in Fig.\ 2 by FFT are demonstrated in Fig.\ 4.
The sharp fundamental peaks indicate 
that $H_{\rm p}$ is uniform over the wide fields.
The small peaks corresponding to the second harmonic waves are observed
because the oscillations are distorted slightly from sinusoidal.
The positions of the peaks, which are inversely proportional to $H_{\rm p}$,
are plotted as a function of the width ($w$) of the junctions
in the inset of Fig.\ 4.
We see that the $1/H_{\rm p}$ is proportional to the width ($w$).

What is the origin of these oscillations?
It should be checked on the relation 
between $H_{\rm p}$ and the number of vortices in the junctions.
Supposing that the Josephson vortices uniformly penetrate into every junction,
the increment of the field ($H_{\rm 0}$) 
which is needed to add one vortex per a junction is
$\phi_0 / w s $, where 
$\phi_0$ ($\simeq 2.07 \times 10^{-7}$ G cm$^{2}$) is flux quantum 
and $s$ (=15\AA) the junction thickness, namely, 
the distance between superconducting CuO$_2$ layers.
$1/H_{\rm 0}(w)$ is shown as a broken line in the inset of Fig.4. 
Interestingly, the experimental values of $1/H_{\rm p}$ are close to 
the twice of $1/H_{\rm 0}(w)$ as shown by a solid line, 
which means that $H_{\rm p}$ corresponds to 
adding ``one'' flux quantum per ``two'' IJJ's. 
We can check 
whether the resistance takes a maximum or a minimum
when the field is $n H_{\rm p}$ ($n$ : integer).
Experimentally, we have shown
that the oscillations have minima at $n H_{\rm p}$.

The above experimental results can be explained by assuming that;
(1) the lattice structure of the Josephson-vortex system is 
the triangular lattice, and 
(2) at both right and left surfaces of the junction,
a surface barrier, i.e., Bean-Livingston barrier \cite{Bean64} 
works on vortices, 
although both attractive image force and repulsive force by Meissner current
are quite small because of a little inhomogeneity of the magnetic field
around vortices in junctions.
For example,
let's consider that the field is 
$n (H_{\rm 0}/2) \approx n H_{\rm p}$.
The Josephson vortices are distributed as schematically shown in Fig.5(a).
This ordered state fits to the width of junctions.
When the vortex lattice flow from left to right,
vortices at right side exit beyond small surface potentials and
vortices would enter at left side simultaneously.
Now,
the total potential which the lattice feels at both sides becomes maximum.
At this situation, the averaged velocity and hence resistance of vortex flow 
has minimum.
In a short time scale, 
the velocity of the vortex lattice would oscillate with time
whenever vortices go over the boundaries.
As the field is increased, additional vortices are forced 
to enter into the ordered lattice.
The Josephson vortex lattice could move more easily 
because of the incommensurability
between the width of the junctions and the lattice spacing of vortices. 
Then, the flow resistance becomes larger than that in the above fitted case. 
When the field goes over $(n+1/2)(H_{\rm 0}/2)$, 
the lattice starts to return to the fitted state. 
At $(n+1)(H_{\rm 0}/2)$, matching between the Josephson vortex lattice 
and the size of junction recovers as shown in Fig.5(b), 
where the Josephson vortices enter in alternate layers comparing with Fig.5(a).
Hence, it can be plausible that 
the oscillation period of the flow resistance is a half of $H_{\rm 0}$. 
When the Lorentz force from the applied current becomes
large enough comparing with the surface potential,
it is expected that the oscillations smear out in this scenario,
which can explain the $I$-$V$ characteristics qualitatively.
More complicated models are also possible, 
which may explain the details of flow dynamics.

There are several theoretical studies with respect to 
the thermodynamic phase diagram of the Josephson vortex system 
in anisotropic layered superconductors \cite{Ivlev90,Balents95,Hu2000}. 
Ivlev et al. \cite{Ivlev90} 
have derived the structure of Josephson-vortex lattice 
as the compressed hexagons of triangular lattice pointing 
along the $c$-axis in high enough fields by using London theory.
However, in the case of strong magnetic fields 
of order $\phi_0/\gamma s^2$, where $\gamma$ is anisotropy ratio,
the three-dimensional(3D) anisotropic London or Ginzburg-Landau approach
is not adequate \cite{Bulaevskii91}.
This field is estimated as 46\ kOe in $\gamma\sim200$ \cite{Iye92}.
It is delicate whether the London theory can be applied at the field
where the oscillations occur.
On the other hand, the extensive Monte Carlo simulations 
with the 3D anisotropic frustrated XY model by Hu {\it et al.} 
have also shown that a triangular lattice is the ground state 
where the Josephson vortices distribute in every layer 
for the anisotropy parameter $\gamma \geq$8
when the filling factor is 1/32 \cite{Hu2000}. 
For $\gamma=$150 as is the case of BSCCO, 
it is known that the compressed triangular lattice is achieved 
even below the tricritical point value ($\sim$ 15 kOe) 
by rescaling \cite{Hu2000}.
These theoretical results are consistent with our experimental ones. 
For the magnetic fields lower than 7kOe, the oscillations 
cannot be observed in all samples 
in spite of the existence of finite flow resistance. 
A reason why the oscillations smear out at the low magnetic fields 
may be that the distance between the vortices is not close 
enough to form a triangular lattice. 
Ivlev {\it et al.} have predicted that a triangular array 
becomes unstable with respect to the formation of 
several new types of vortex lattices in the low fields \cite{Ivlev90}. 
In addition to the ground state, 
the existence of several metastable states has been indicated \cite{Ivlev90}. 
In this situation, the periodic oscillations could not be expected.

Similar phenomena have been reported 
in artificially structured materials as described above
\cite{Pruymboom88,Gurevich96},
although the origins are different from our case.
Recently, 
in YBCO single crystals of 60\ K phase 
which is more anisotropic than 90\ K phase,
an oscillatory behavior of the in-plane resistivity 
in the field parallel to the $ab$ plane
have been reported \cite{Gordeev2000}.
In their experiments,
the oscillations have minima when the period of the layered structure of YBCO
matches the intervortex spacing in the direction normal to the layers.
The period of the oscillation is proportional to $B^{-0.5}$ and 
is independent of the size of samples.
This phenomenon originates from the commensurability 
between $s$ and vortex-lattice space along the $c$-axis,
while the two lengths, the junction width $w$ 
and vortex-lattice spacing along the $ab$ plane are important 
in present case.

In summary,
we have measured the flow resistance of Josephson vortices in
IJJ's of BSCCO single crystals.
The novel oscillations of the resistance as a function of field 
by the Josephson-vortex flow have been found.
The period of the oscillations depends only on the width of junctions.
The fact that the period is $H_{\rm 0}/2$, 
which is needed to add ``{\it one}'' Josephson vortex per``{\it two}'' layers,
suggests that the lattice structure of the Josephson vortices 
is triangular.
The oscillations originate from commensuration
between the triangular lattice and the width of the junction
perpendicular to the field.
At present, our measurements do not access the boundary 
between the lattice and liquid phase of Josephson-vortex system. 
In the future, further improvements of measurements would make 
it possible to investigate the phase diagram of 
Josephson-vortex system including a liquid state by using this phenomenon.

The authors thank X.~Hu and M.~Machida for useful discussions,
N.~Ishikawa for technical supports of FIB.



\begin{figure}[htbp]
\begin{center}
\includegraphics[width=0.8\linewidth]{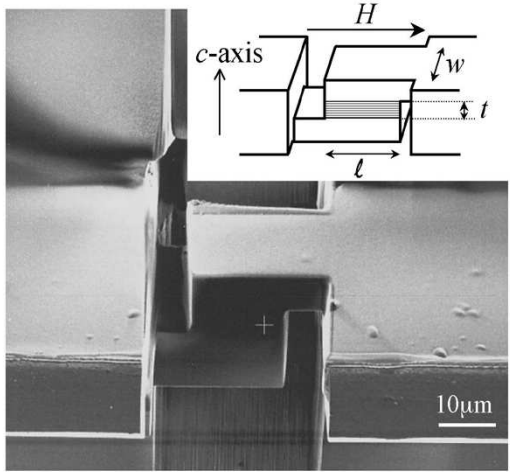}
\end{center}
\caption{
An FIB image of a typical sample. 
Schematic drawing of the junction (hatched part) is shown in the inset. 
Magnetic field is applied to the horizontal direction. 
Definition of the junction dimensions is also indicated.
}
\end{figure}

\begin{figure}[t]
\begin{center}
\includegraphics[width=0.9\linewidth]{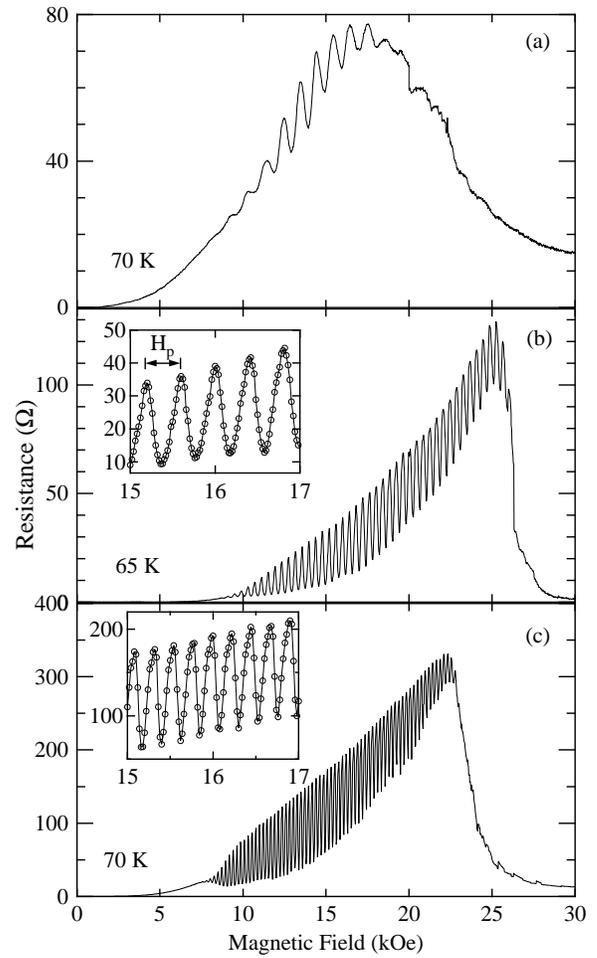}
\end{center}
\caption{
Flow resistance of Josephson vortices 
as a function of magnetic field measured on the sample A (a),
B (b), and C (c) with the applied currents of
1, 1, and 10 $\mu$A, respectively.
The enlarged figures of the flow-resistance oscillations of the sample B and C
are shown in the insets of (b) and (c), respectively.
}
\end{figure}

\begin{figure}[t]
\begin{center}
\includegraphics[width=0.8\linewidth]{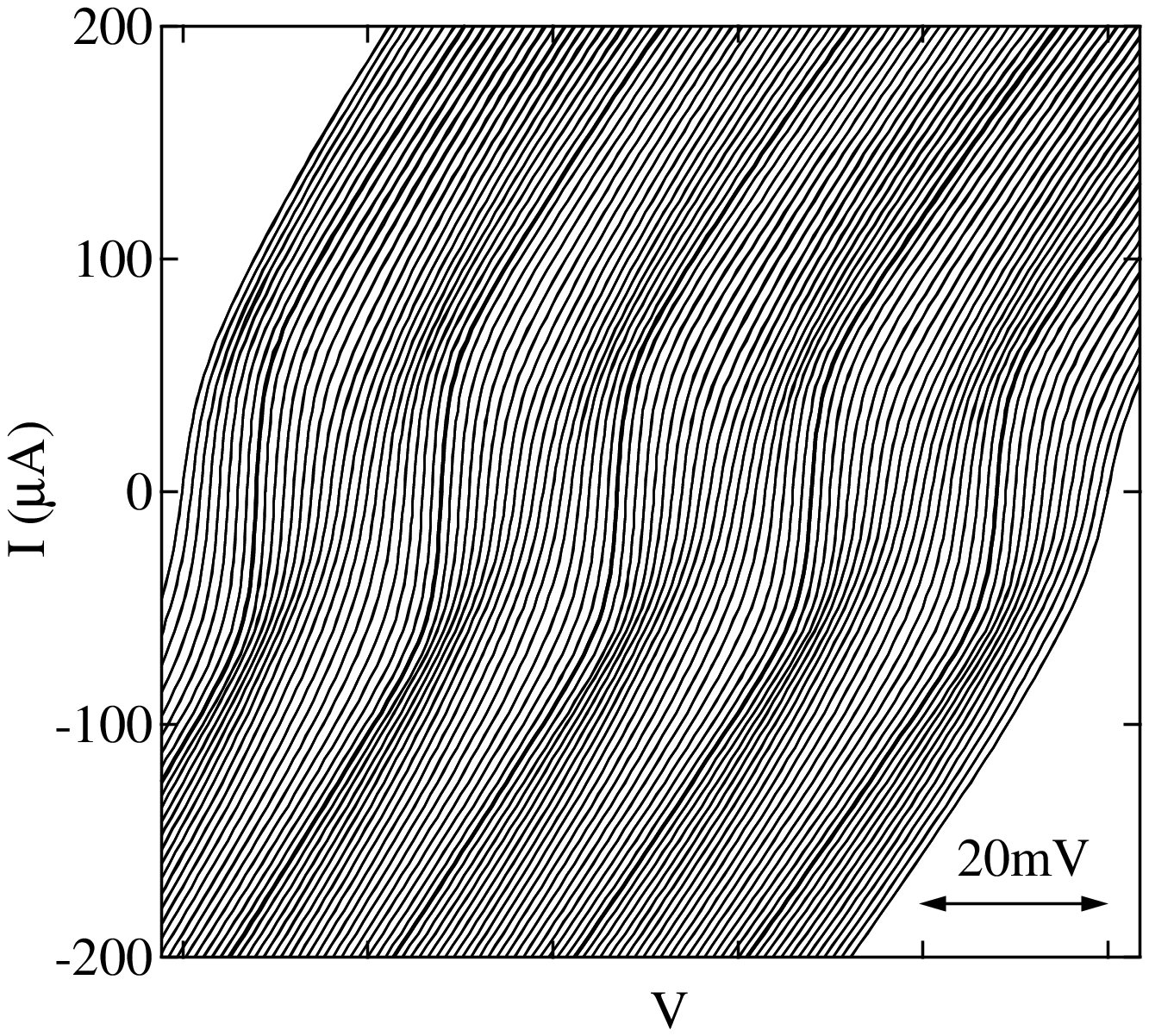}
\end{center}
\caption{
$I$-$V$ characteristics of the sample B
at various fields from 14\ k to 16\ kOe at every 20\ Oe in 65\ K.
Each curve is shifted by 1mV step.
Bold lines show $I$-$V$ curves where the resistance is minimum.
}
\end{figure}

\begin{figure}[htbp]
\begin{center}
\includegraphics[width=0.8\linewidth]{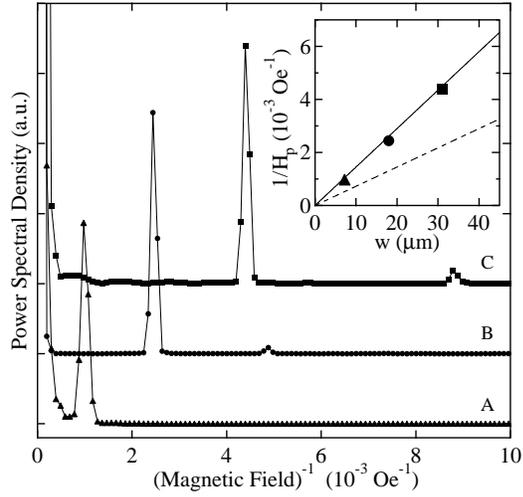}
\end{center}
\caption{
Power spectral densities of the oscillations for sample A, B and C. 
Vertical axes are shifted properly. 
In the inset, solid circles show the inverse of $H_{\rm p}$, 
which are estimated as the positions of the peaks in spectra, 
as a function of width ($w$). 
The broken and solid lines represent $1/H_{\rm 0}(w)$ and its twice, 
respectively.
}
\end{figure}

\begin{figure}[htbp]
\begin{center}
\includegraphics[width=213bp, height=79bp]{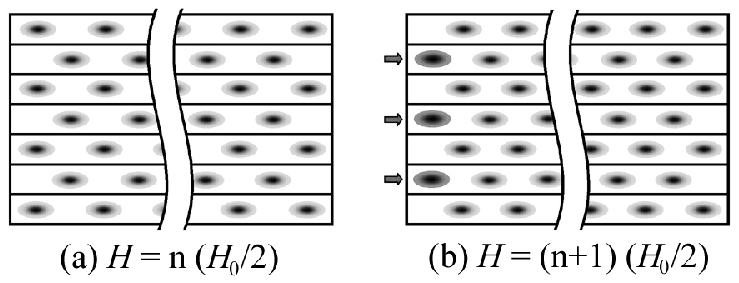}
\end{center}
\caption{
Schematic pictures of the configuration of Josephson-vortex lattice 
when the magnetic fields are reached to 
(a) $n(H_{\rm 0}/2)$ and (b) $(n+1)(H_{\rm 0}/2)$, 
where $n$ is integer.}
\end{figure}


\begin{references}

\bibitem{Kleiner92}
R.~Kleiner, F.~Steinmeyer, G.~Kunkel, and P.~M\"{u}ller, 
Phys. Rev. Lett. {\bf 68} (1992) 2394.

\bibitem{Tachiki89}
M.~Tachiki, and S.~Takahashi,
Solid State Commun. {\bf 70} (1989) 291.

\bibitem{Ivlev90}
B.I.~Ivlev, N.B.~Kopnin, and V.L.~Pokrovsky,
J. Low. Temp. Phys. {\bf 80} (1990) 187.

\bibitem{Balents95}
L.~Balents, and D.R.~Nelson,
Phys. Rev. B {\bf 52} (1995) 12951.

\bibitem{Hu2000}
X.~Hu, and M.~Tachiki,
Phys. Rev. Lett. {\bf 85} (2000) 2577.

\bibitem{Kwok94}
W.K.~Kwok, J.~Fendrich, U.~Welp, S.~Fleshler, J.~Downey, and G.W.~Crabtree,
Phys. Rev. Lett. {\bf 72} (1994) 1088.

\bibitem{Schilling97}
A.~Schilling, R.A.~Fisher, N.E.~Phillips, U.~Welp, W.K.~Kwok, and G.W.~Crabtree,Phys. Rev. Lett. {\bf 78} (1997) 4833.

\bibitem{Ishida97}
T.~Ishida, K.~Okuda, A.I.~Rykov, S.~Tajima, and I.~Terasaki,
Phys. Rev. B {\bf 58} (1998) 5222.

\bibitem{Gordeev2000}
S.N.~Gordeev, A.A.~Zhukov, P.A.J.~de Groot, A.G.M.~Jansen, R.~Gagnon, and
L.~Taillefer,
Phys. Rev. Lett. {\bf 85} (2000) 4594.

\bibitem{Pruymboom88}
A.~Pruymboom, P.H.~Kes, E. van der Drift and S.~Radelaar,
Phys. Rev. Lett. {\bf 60} (1988) 1430.

\bibitem{Gurevich96}
A.~Gurevich, E.~Kadyrov, and D.C.~Larbalestier, 
Phys. Rev. Lett. {\bf 77} (1996) 4078.

\bibitem{Mochiku97}
T.~Mochiku, K.~Hirata, and K.~kadowaki, 
Physica C {\bf 282-287}, 475 (1997).

\bibitem{Kim99}
S.-J.~Kim, Yu.~I. Latyshev, and T.~Yamashita, 
Appl. Phys. Lett. {\bf 74} (1999) 1156.


\bibitem{Lee95}
J.U.~Lee, J.E.~Nordman, and G.~Hohenwarter,
Appl. Phys. Lett. {\bf 67} (1995) 1471.

\bibitem{Hechtfischer97}
G.~Hechtfischer, R.~Kleiner, K.~Schlenga, W.~Walkenhorst, P.~M\"{u}ller,
and H.L.~Johnson,
Phys. Rev. B {\bf 55} (1997) 14638.

\bibitem{Yurgens99}
A.~Yurgens, D.~Winkler, T.~Claeson, G.~Yang, I.F.G.~Parker, and
C.E.~Gough,
Phys. Rev. B {\bf 59} (1999) 7196.

\bibitem{Bean64}
C.P.~Bean and J.D.~Livingston, Phys. Rev. Lett. {\bf 12} (1964) 14.

\bibitem{Bulaevskii91}
L.~Bulaevskii and J.R.~Clem, Phys. Rev. B {\bf 44} (1991) 10234.

\bibitem{Iye92}
Y.~Iye, I.~Oguro, and T.~Tamegai, Physica C {\bf 199} (1992) 154.


\end{references}
\end{document}